# Hadronic Coupling Constants in Lattice QCD

R.L. Altmeyer[a], M. Göckeler[b], R. Horsley[b], E. Laermann[c],

G. Schierholz[a,b] and P.M. Zerwas[a]

[a] Deutsches Elektronen-Synchrotron DESY,
D-22603 Hamburg, Germany

[b] Gruppe Theorie der Elementarteilchen,
Höchstleistungsrechenzentrum HLRZ, c/o Forschungszentrum Jülich,
D-52425 Jülich, Germany

[c] Fakultät für Physik, Universität Bielefeld,
D-33501 Bielefeld, Germany

## Abstract

We calculate the hadronic coupling constants $g_{NN\pi}$ and $g_{\rho\pi\pi}$ in QCD, including dynamical quarks in the framework of staggered fermions in the lattice approach. For the nucleon–pion coupling we obtain $g_{NN\pi} = 13.8 \pm 5.8$, to be compared with the experimental value $13.13 \pm 0.07$ [1]. The $\rho\pi\pi$ coupling has been analysed for two different sets of operators with the averaged result $g_{\rho\pi\pi} = 4.2 \pm 1.9$ which is to be compared with the experimental value $6.06 \pm 0.01$ [2].

# 1 Hadronic coupling constants in lattice QCD

Recent lattice QCD calculations have given encouraging results for the hadronic mass spectrum. This success motivates the launching of more complicated analyses, in particular the calculation of dynamical properties of hadrons. The aim of the present analysis is the calculation of coupling constants which parameterise the strength of the interaction between hadrons at low energies, above all the classical nucleon-pion coupling constant.

We have based the analysis on the staggered fermion action with four degenerate flavours of dynamical fermions. The lattice size is $L^3 \times T = 16^3 \times 24$ and we have generated 85 configurations at $\beta = 5.35$, $m = 0.01$ using the hybrid Monte Carlo algorithm. Details with respect to the production of configurations can be recalled from ref.[3].

Hadronic coupling constants can be determined by calculating 2– and 3–point functions on the lattice using the path integral method. Compared to the analysis of the mass spectrum, however, the calculation of coupling constants is, from the theoretical as well as from the numerical point of view, much more involved.

Above all we have to choose a set of operators very carefully so that the following criteria are fullfilled: First the operators should have a good overlap with the hadronic particle state under consideration and, if possible, a high suppression of the unwanted parity partner state created by the usual quasi–local staggered operators. Secondly, because the continuum limit of the staggered lattice action corresponds to a theory describing four mass degenerate quark flavours the operators have to be chosen such that the coupling is allowed by $SU(4)_F$ flavour symmetry group selection rules.

Next one has to find a function to parameterise the 3–point function in which the hadronic coupling constant appears as a parameter. Besides the usual transfer matrix techniques and the LSZ formalism, which relates the correlation functions to coupling constants, one has to determine the corresponding $SU(4)_F$ Clebsch–Gordan (CG) coefficients because the coupling constants are defined as flavour independent numbers. This, however, requires in a first step to relate the lattice operators to the corresponding operators in the continuum so that $SU(4)_F$ quantum numbers can be assigned to the states.

Due to the technically complicated analysis, we proceed in two steps and consider the $g_{\rho\pi\pi}$ and $g_{NN\pi}$ coupling constants separately (see also ref.[4] and [5]).

# 2 The $g_{\rho\pi\pi}$ coupling constant

The effective interaction Lagrangian which parameterises the decay of a $\rho$ meson into two pions is of the form

$$\mathcal{L}_{int}^{eff}(x) = g_{\rho\pi\pi} c_{abc} \rho_\nu^a(x) [\pi^b(x) \overleftrightarrow{\partial}^\nu \pi^c(x)]. \quad (1)$$

Here $c_{abc}$ is the $SU(4)_F$ Clebsch–Gordan coefficient which ensures that the Lagrangian behaves like a scalar under flavour transformations; $g_{\rho\pi\pi}$ does not depend on flavour quantum numbers.

For the $\rho$ meson we used the local operator [No.4][1] because it delivered quite stable results in the analysis of the mass spectrum . Moreover, the numerical evaluation of the 3–

---
[1] The numbers of the operators refer to the numbers used in ref.[3].

point function is much eased by using local operators. However, like most of the staggered hadron operators the $\rho$ operator has the unwanted property of coupling also to a state with opposite parity, namely a $b_1$ meson. This parity-partner state contributes to the correlation functions so that additional terms have to be taken into account, thus leading to more complicated parameterisations. Therefore we used also a non–local–in–time (NLT) $\rho$ operator [No.23] which belongs to the same representation of the lattice symmetry group as the local $\rho$ but whose coupling to the parity partner meson is dynamically suppressed.

One of the pion operators ($\pi_b$) was chosen to be the lattice Goldstone (LG) pion [No.2] which, besides locality, has the additional advantage of being the lightest pionic state in our simulation[2] so that the kinematics of the $\rho$ decay is closest to the situation in the continuum. Moreover, the parity partner of the pion created by this operator corresponds to an exotic state and is highly suppressed so that it can be neglected in the analysis.

With these operators given, the third operator ($\pi_c$) has to be chosen in a way that the CG coefficient $c_{abc}$ is non–zero. Therefore we adopted a non–local pion, the 1–link local–in–time (LT) operator [No.7] which has an exotic, suppressed parity partner. We then calculate the 3-point correlation function

$$\Gamma_3(t_1, t_2) \stackrel{\mathrm{def}}{=} \epsilon_\mu(\vec{q}_\rho) \langle \tilde{\rho}_a^\mu(\vec{q}_\rho, t_1) \tilde{\pi}_b(\vec{q}_\pi, t_2) \pi_c(0) \rangle, \tag{2}$$

$\epsilon_\mu$ being the polarisation vector of the $\rho$ meson. The tilde indicates that the corresponding field operator is defined in momentum space, e.g.

$$\tilde{\pi}(\vec{q}, t) = \frac{1}{V} \sum_{\vec{x} \in V} e^{-i\vec{q}\vec{x}} \pi(\vec{x}, t), \tag{3}$$

where $V$ is the spatial volume of the lattice.

The time slice $t_2$ of the pion has been fixed to $t_2 = 4$ and we evaluated the correlation function for all times $t_1$ of the $\rho$ meson. The $\rho$ meson was given the minimal lattice momentum $\vec{q}_\rho = \frac{2\pi}{L} \vec{e}_z$ in the $z$ direction, and $\vec{q}_\pi$ was set to $\vec{0}$. The pion source $\pi$ then projects on a state with non-vanishing momentum.

To define a parameterisation of this correlation function we insert complete sets of states in eq.(2). Assuming large time distances between the operators we can restrict ourselves to the low lying particle states and the following expression can then be derived for $T \gg t_1 \gg t_2 \gg 0$:

$$\begin{aligned}
\Gamma_3(t_1, t_2) &\approx \langle \rho_a, \vec{q}_\rho | \tilde{\pi}_b(\vec{0}) | \pi_c, \vec{q}_\rho \rangle \sqrt{Z_{\rho_a} Z_{\pi_c}} e^{-E_{\rho_a}(t_1-t_2)} e^{-E_{\pi_c} t_2} (-1)^{t_1+t_2} + \\
&\quad \langle \pi_c, \vec{q}_\rho | \tilde{\rho}_a(\vec{q}_\rho) | \pi_b, \vec{0} \rangle \sqrt{Z_{\pi_c} Z_{\pi_b}} e^{-E_{\pi_b}(t_1-t_2)} e^{-E_{\pi_c}(T-t_1)} (-1)^{t_1+t_2} + \\
&\quad \langle \pi_b, \vec{0} | \tilde{\pi}_c(\vec{q}_\rho) | \rho_a, \vec{q}_\rho \rangle \sqrt{Z_{\pi_b} Z_{\rho_a}} e^{-E_{\rho_a}(T-t_1)} e^{-E_{\pi_b} t_2} (-1)^{t_1+t_2} + \\
&\quad \langle \pi_b, \vec{0} | \tilde{\pi}_c(\vec{q}_\rho) | b_{1a}, \vec{q}_\rho \rangle \sqrt{Z_{\pi_b} Z_{b_{1a}}} e^{-E_{b_{1a}}(T-t_1)} e^{-E_{\pi_b} t_2} (-1)^{t_2} + \\
&\quad \langle b_{1a}, \vec{q}_\rho | \tilde{\pi}_b(\vec{0}) | \pi_c, \vec{q}_\rho \rangle \sqrt{Z_{b_{1a}} Z_{\pi_c}} e^{-E_{b_{1a}}(t_1-t_2)} e^{-E_{\pi_c} t_2}.
\end{aligned} \tag{4}$$

---

[2]This is due to the fact that our lattice is not close enough to the continuum to completely restore the flavour symmetry in the chiral sector.

$Z_{\mathcal{M}}$ denotes the 1-meson amplitude of the operator $\mathcal{M}$, $\sqrt{Z_{\mathcal{M}}} = \langle 0|\mathcal{M}|\,\text{Meson}\,\rangle$. These constants together with the masses and energies of the different particle states are taken from the analysis of 2–point functions. The corresponding formula for $T \gg t_2 \gg t_1 \gg 0$ is

$$\begin{aligned}
\Gamma_3(t_1, t_2) &\approx \langle \pi_b, \vec{0}|\tilde{\rho}_a(\vec{q}_\rho)|\pi_c, \vec{q}_\rho\rangle \sqrt{Z_{\pi_c} Z_{\pi_b}} e^{-E_{\pi_b}(t_2-t_1)} e^{-E_{\pi_c} t_1}(-1)^{t_1+t_2} + \\
&\quad \langle \rho_a, \vec{q}_\rho|\tilde{\pi}_c(\vec{q}_\rho)|\pi_b, \vec{0}\rangle \sqrt{Z_{\pi_b} Z_{\rho_a}} e^{-E_{\rho_a} t_1} e^{-E_{\pi_b}(T-t_2)}(-1)^{t_1+t_2} + \\
&\quad \langle \pi_c, \vec{q}_\rho|\tilde{\pi}_b(\vec{0})|\rho_a, \vec{q}_\rho\rangle \sqrt{Z_{\rho_a} Z_{\pi_c}} e^{-E_{\rho_a}(t_2-t_1)} e^{-E_{\pi_c}(T-t_2)}(-1)^{t_1+t_2} + \\
&\quad \langle b_{1a}, \vec{q}_\rho|\tilde{\pi}_c(\vec{q}_\rho)|\pi_b, \vec{0}\rangle \sqrt{Z_{\pi_b} Z_{b_{1a}}} e^{-E_{b_{1a}} t_1} e^{-E_{\pi_b}(T-t_2)}(-1)^{t_2} + \\
&\quad \langle \pi_c, \vec{q}_\rho|\tilde{\pi}_b(\vec{0})|b_{1a}, \vec{q}_\rho\rangle \sqrt{Z_{b_{1a}} Z_{\pi_c}} e^{-E_{b_{1a}}(t_2-t_1)} e^{-E_{\pi_c}(T-t_2)}.
\end{aligned} \tag{5}$$

In principle eq.(4) should include terms accounting for the parity partners of the mesons. However, as the opposite parity states to both pions are exotic and therefore suppressed (i.e. $Z \approx 0$, as can be seen in the analysis of 2–point functions), they need not be taken into account. On the other hand, the coupling described by the matrix element where the $\rho$ meson parity partner and the two pion states are involved, violates lattice symmetry and is therefore forbidden. For the NLT $\rho$ also the last two terms in eq.(4) can be neglected as the the coupling of this operator to the $b_1$ state is suppressed.

In order to find the relation between the matrix elements, e.g. $\langle \rho_a|\tilde{\pi}_b|\pi_c\rangle$, and the coupling constants we apply the LSZ reduction formula and evaluate the resulting Green's function according to the effective interaction Lagrangian in eq.(1). The resulting expression consists of a free propagator for the $\pi_b$ particle, a kinematic factor $K(\vec{q}_\rho)$ for the truncated vertex and the hadronic coupling constant:

$$\langle \rho_a, \vec{q}_\rho|\tilde{\pi}_b(\vec{0})|\pi_c, \vec{q}_\rho\rangle = g_{\rho\pi\pi} c_{abc} K(\vec{q}_\rho) \sqrt{Z_{\pi_b}} P((E_{\rho_a} - E_{\pi_c}), E_{\pi_b}). \tag{6}$$

$P(E, \omega)$ denotes the free lattice particle propagator for a particle with pole energy $\omega$ propagating at energy $E$ which is defined through

$$P(E, \omega) = \frac{\sinh \omega}{\cosh \omega - \cosh E}. \tag{7}$$

The kinematic factor for the process $\rho \to \pi\pi$ is given by

$$K(\vec{q}_\rho) = |\vec{q}_\rho| \frac{m_{\pi_b} + m_{\pi_c}}{m_{\rho_a}}. \tag{8}$$

The parameter function was fitted to the numerical data by a procedure minimizing $\chi^2$. The masses and energies of the various states could be kept fixed as they are already known from the analysis of 2–point functions. Thus we have one fit parameter for the correlation function with the NLT $\rho$ operator and 3 fit parameters when using the LT $\rho$ operator. However, in the latter case only one parameter, namely $g_{\rho\pi\pi}$ is of physical interest whereas the others are proportional to the matrix elements where the pion operator serves as an interpolating field between a $b_1$ and a $\pi$ state. The time interval was varied in the fit procedure; however the variation could be done only in a restricted range since due to the involved parameter function the fit soon becomes unstable when the number of fitting points is reduced. The result of the fit is a product of the coupling constant $g_{\rho\pi\pi}$, the amplitudes $Z$ and the CG coefficients. The $Z$ factors, like the masses, are known from 2–point functions and can readily be divided out.

To derive the value for $g_{\rho\pi\pi}$, we still have to determine the CG coefficients $c_{abc}$. The flavour non–singlet meson states are members of an $SU(4)_F$ 15-plet so that we need to calculate CG coefficients for the product of representations $15 \otimes 15 \to 15$. The CG coefficients have already been calculated for this case in ref.[6], but we still have to relate the lattice operators to continuum operators. To do so, we write the staggered mesonic operators in the following form

$$M_{SF}(y) = \overline{q}(y)[\Gamma_S \otimes \Gamma_F]q(y), \qquad (9)$$

where $\Gamma_S$, $\Gamma_F$ define the spin and flavour content of the meson and $q$ denotes a staggered quark field defined on a hypercube of the lattice:

$$q^{\alpha a}(y) \stackrel{\text{def}}{=} \sum_\eta \Gamma_\eta^{\alpha a} \chi(2y + \eta). \qquad (10)$$

$\Gamma_\eta^{\alpha a} = (\gamma_1^{\eta_1} \cdots \gamma_4^{\eta_4})^{\alpha a}$ is a product of $4\times 4$ $\gamma$ matrices and $\alpha, a$ denote spin and flavour indices of the quark, respectively. To assign continuum flavour values to the lattice operators we call the $a = 1$ component of the lattice quark operator a $u$ quark, the $a = 2$ component a $d$ quark, the $a = 3$ component an $s$ quark and the $a = 4$ a $c$ quark.[3] If we express the flavour matrix $\Gamma_F$ in eq.(9) through $SU(4)_F$ Gell–Mann matrices, we are able to identify the lattice meson operators with continuum operators used in ref.[6]. On the other hand, the relationship between the lattice meson operator in eq.(9) and the lattice operators given in ref.[3] can be readily found by using the above formulae and trace theorems for $\gamma$ matrices.

As the $\Gamma_F$ matrices correspond to linear combinations of $SU(4)_F$ Gell–Mann matrices the staggered meson operators correspond to linear combinations of continuum operators

$$M_a^{Latt} \leftrightarrow \sum_b (f_M)_a^b M_b^{Cont}. \qquad (11)$$

$M_a^{Latt}$ denotes a member of the lattice meson multiplet and $M_b^{Cont}$ is a member of the continuum multiplet. Thus the lattice coefficient $c_{abc}$ in eq.(6) is given by

$$c_{abc} = \sum_{d,e,f} (f_{\rho_a})_a^d (f_{\pi_b})_b^e (f_{\pi_c})_c^f c_{def}^{Cont}, \qquad (12)$$

where the continuum CG coefficients can be taken from ref.[6]. To be specific, we give the correspondence of the lattice meson operators, relevant in the evaluation of the $g_{\rho\pi\pi}$

---

[3]This assignment is arbitrary as all staggered quark flavours are mass degenerate. However, consistent definitions have to be used when group theoretical relations including different operators are evaluated.

coupling constant, to the continuum meson operators:

$$\rho_a^{Latt} \leftrightarrow \frac{1}{2}(K^{*+} - K^{*-} + D^{*+} - D^{*-}) \quad (13)$$

$$\pi_b^{Latt} \leftrightarrow \frac{1}{\sqrt{2}}(\frac{2}{\sqrt{3}}\eta_8 + \frac{2}{\sqrt{6}}\eta_{15})$$

$$\pi_c^{Latt} \leftrightarrow \frac{1}{2}(K^+ + K^- - D^+ - D^-).$$

These relations determine the matrices $(f_M)_a^b$ and thus allow the derivation of the CG coefficient $c_{abc}$,

$$c_{abc} = \sqrt{2}. \quad (14)$$

Now we are ready to determine the $g_{\rho\pi\pi}$ coupling constant from the numerical data which are shown in fig.1 for the case of the local $\rho$ operator. We find for the two different cases of the local and the NLT $\rho$ operator

$$\begin{aligned} g_{\rho\pi\pi} &= 3.6 \pm 2.5 \quad \text{(LT)}, \\ g_{\rho\pi\pi} &= 4.7 \pm 2.8 \quad \text{(NLT)}, \end{aligned} \quad (15)$$

from which we derive the average value

$$g_{\rho\pi\pi} = 4.2 \pm 1.9. \quad (16)$$

Compared to the experimental value of $g_{\rho\pi\pi} = 6.06 \pm 0.01$ our results are too small but still compatible with the measurement within $1\sigma$ of the large errors. Moreover, our results are larger than the results in ref.[7] based on Wilson fermions for a smaller lattice which spread between $g_{\rho\pi\pi} \approx 2.8$ and $g_{\rho\pi\pi} \approx 3$.

## 3 The $g_{NN\pi}$ coupling constant

The effective Lagrangian associated with the $NN\pi$ coupling is defined as

$$\mathcal{L}_{int}^{eff}(x) = g_{NN\pi} c'_{abc} \overline{N}^a(x)\gamma_5 N^b(x)\pi^c(x). \quad (17)$$

This equation holds only for $SU(2)_F$ flavour symmetry subgroup or, more precisely, if the nucleon belongs to a 2-dimensional representation and the meson to a 3-dimensional representation of the same $SU(2)_F$ subgroup of the full flavour symmetry group. For the full $SU(4)_F$ flavour symmetry in a general case, however, we have to consider two independent coupling constants $g_1$ and $g_2$ so that eq.(17) must be interpreted as

$$\mathcal{L}_{int}^{eff}(x) = (g_1 c_{abc}^1 + g_2 c_{abc}^2)\overline{N}^a(x)\gamma_5 N^b(x)\pi^c(x). \quad (18)$$

Mathematically, this is due to the fact that the product of the 20–dimensional representation to which the baryon belongs in $SU(4)_F$, and the 15–dimensional meson representation reduces, among others, to two different 20-dimensional baryon representations. This situation is analogous to $SU(3)_F$ where $F$ and $D$ couplings, $g_F^A$ and a $g_D^A$, must properly be distinguished.

The set of operators we chose in this case consists of a local polarised baryon operator, the lattice Goldstone pion [No.2] and a baryonic wall source operator. The baryon should be polarised since the contribution of the $NN\pi$ coupling to the correlation function vanishes otherwise. The Goldstone pion again was used because of its low mass, its local character and the suppressed opposite parity state.

We decided to use a wall source operator for the second baryon as it has a much better overlap with the ground state particle compared to the local operator. This is especially important for the case of 3–point functions because time distances between the operators are of course even smaller if three operators are defined on the lattice so that the suppression of higher states by the exponential functions is worse. Indeed, the quality of the numerical data improved significantly compared to earlier attempts where we used a local source. Moreover, the results of the analysis of 2–point functions, especially the amplitude $Z_N$, are much more stable when the wall source operator is employed.

The 3-point function we evaluated numerically, is of the form

$$\Gamma_3(t_1, t_2) = \left\langle \widetilde{N}_a(\vec{p}, t_1) \widetilde{\pi}_b(-\vec{p}, t_2) \overline{N}_c^W \right\rangle, \tag{19}$$

where $N$ denotes a nucleon operator with positive polarisation in the $z$ direction and $N^W$ is the nucleon wall source operator. Since the wall source operator we used projects onto states with vanishing momentum we have to give opposite momenta to the nucleon sink and the meson operator (the $NN\pi$ coupling vanishes as $\vec{p} \to 0$). $\vec{p}$ was chosen to be the minimal lattice momentum in the $z$ direction. The time of the sink nucleon was fixed at $t_1 = 8$ and the correlation function has been evaluated for all time slices $t_2$ of the pion.

A parameterisation of this function can again be found by using transfer matrix techniques and inserting a complete set of particle states between the operators. Neglecting, for large time distances between the operators, the influence of higher mass states, we find the following parameterisation for $T \gg t_2 \gg t_1 \gg 0$:

$$\left\langle \widetilde{\pi}_b(-\vec{p}, t_2) \widetilde{N}_a(\vec{p}, t_1) \overline{N}_c^W(0) \right\rangle \approx \tag{20}$$

$$\left\langle \pi_b, -\vec{p} \left| \widetilde{N}_a(\vec{p}) \right| N_c, \vec{0} \right\rangle \sqrt{Z_{\pi_b} Z_{N_c}} e^{-E_{\pi_b}(t_2-t_1)} e^{-E_{N_c} t_1} (-1)^{(t_2+t_1)} +$$

$$\left\langle N_a, \vec{p} \left| \overline{N}_c^W(\vec{0}) \right| \pi_b, \vec{p} \right\rangle \sqrt{Z_{\pi_b} Z_{N_a}} e^{-E_{\pi_b}(T-t_2)} e^{-E_{N_a} t_1} (-1)^{t_2} \xi_C^\pi +$$

$$\left\langle \overline{N}_a, \vec{p} \left| \widetilde{\overline{\pi}}_b(-\vec{p}) \right| \overline{N}_c, \vec{0} \right\rangle \sqrt{Z_{N_a} Z_{N_c}} e^{-E_{N_c}(T-t_2)} e^{-E_{N_a}(t_2-t_1)} (-1)^{t_1} \xi_C^\pi \sigma_C +$$

$$\left\langle \pi_b, -\vec{p} \left| \widetilde{N}_a(\vec{p}) \right| N_c', \vec{0} \right\rangle \sqrt{Z_{\pi_b} Z_{N_c'}} e^{-E_{\pi_b}(t_2-t_1)} e^{-E_{N_c'} t_1} (-1)^{t_2} +$$

$$\left\langle N_a', \vec{p} \left| \overline{N}_c^W(\vec{0}) \right| \pi_b, -\vec{p} \right\rangle \sqrt{Z_{\pi_b} Z_{N_a'}} e^{-E_{\pi_b}(T-t_2)} e^{-E_{N_a'} t_1} (-1)^{(t_2+t_1)} \xi_C^\pi +$$

$$\left\langle \overline{N'}_a, \vec{p} \left| \widetilde{\overline{\pi}}_b(-\vec{p}) \right| \overline{N'}_c, \vec{0} \right\rangle \sqrt{Z_{N_a'} Z_{N_c'}} e^{-E_{N_c'}(T-t_2)} e^{-E_{N_a'}(t_2-t_1)} \xi_C^\pi \sigma_C +$$

$$\left\langle \overline{N'}_c, \vec{0} \left| \widetilde{\overline{\pi}}_b(-\vec{p}) \right| \overline{N}_a, \vec{p} \right\rangle \sqrt{Z_{N_a'} Z_{N_c}} e^{-E_{N_c'}(T-t_2)} e^{-E_{N_a}(t_2-t_1)} (-1)^{(t_2+t_1)} \xi_C^{\pi'} \sigma_C +$$

$$\left\langle \overline{N}_c, \vec{0} \left| \widetilde{\overline{\pi}}_b(-\vec{p}) \right| \overline{N'}_a, \vec{p} \right\rangle \sqrt{Z_{N_a'} Z_{N_c}} e^{-E_{N_c}(T-t_2)} e^{-E_{N_a'}(t_2-t_1)} (-1)^{t_2} \xi_C^{\pi'} \sigma_C,$$

and for $T \gg t_1 \gg t_2 \gg 0$:

$$\left\langle \widetilde{N}_a(\vec{p}, t_1) \widetilde{\pi}_b(-\vec{p}, t_2) \overline{N}_c^W(0) \right\rangle \approx \tag{21}$$

$$\left\langle N_a, \vec{p} \,\middle|\, \widetilde{\pi}_b(-\vec{p}) \,\middle|\, N_c, \vec{0} \right\rangle \sqrt{Z_{N_a} Z_{N_c}} e^{-E_{N_c} t_2} e^{-E_{N_a}(t_1 - t_2)} +$$

$$\left\langle \overline{\pi}_b, \vec{p} \,\middle|\, \overline{N}_c^W(\vec{0}) \,\middle|\, \overline{N}_a, \vec{p} \right\rangle \sqrt{Z_{\pi_b} Z_{N_a}} e^{-E_{\pi_b} t_2} e^{-E_{N_a}(T - t_1)} (-1)^{(t_1 + t_2)} \xi_C^\pi \sigma_C +$$

$$\left\langle \overline{N}_c, \vec{0} \,\middle|\, \widetilde{N}_a(\vec{p}) \,\middle|\, \overline{\pi}_b, -\vec{p} \right\rangle \sqrt{Z_{\pi_b} Z_{N_c}} e^{-E_{\pi_b}(t_1 - t_2)} e^{-E_{N_c}(T - t_1)} (-1)^{t_2} \sigma_C +$$

$$\left\langle N'_a, \vec{p} \,\middle|\, \widetilde{\pi}_b(-\vec{p}) \,\middle|\, N'_c, \vec{0} \right\rangle \sqrt{Z_{N'_a} Z_{N'_c}} e^{-E_{N'_c} t_2} e^{-E_{N'_a}(t_1 - t_2)} (-1)^{t_1} +$$

$$\left\langle \overline{\pi}_b, -\vec{p} \,\middle|\, \overline{N}_c^W(\vec{0}) \,\middle|\, \overline{N'}_a, \vec{p} \right\rangle \sqrt{Z_{\pi_b} Z_{N'_a}} e^{-E_{\pi_b} t_2} e^{-E_{N'_a}(T - t_1)} (-1)^{t_2} \xi_C^\pi \sigma_C +$$

$$\left\langle \overline{N'}_c, \vec{0} \,\middle|\, \widetilde{N}_a(\vec{p}) \,\middle|\, \overline{\pi}_b, -\vec{p} \right\rangle \sqrt{Z_{\pi_b} Z_{N'_c}} e^{-E_{\pi_b}(t_1 - t_2)} e^{-E_{N'_c}(T - t_1)} (-1)^{(t_1 + t_2)} \sigma_C +$$

$$\left\langle N_a, \vec{p} \,\middle|\, \widetilde{\pi}_b(-\vec{p}) \,\middle|\, N'_c, \vec{0} \right\rangle \sqrt{Z_{N'_a} Z_{N_c}} e^{-E_{N'_c} t_2} e^{-E_{N_a}(t_1 - t_2)} (-1)^{t_2} +$$

$$\left\langle N'_a, \vec{p} \,\middle|\, \widetilde{\pi}_b(-\vec{p}) \,\middle|\, N_c, \vec{0} \right\rangle \sqrt{Z_{N'_a} Z_{N_c}} e^{-E_{N_c} t_2} e^{-E_{N'_a}(t_1 - t_2)} (-1)^{(t_1 + t_2)}.$$

$\overline{\pi} = \mathcal{C} \pi \mathcal{C}^{-1}$ denotes the pion operator transformed by lattice charge conjugation and $\xi_C^\pi$ is the corresponding lattice charge conjugation number of the pion. $N'$ is the parity partner state of the nucleon $N$. $\sigma_C$ denotes the eigenvalue of the baryon state under a twofold application of the lattice charge conjugation.

To interpret the matrix elements in the above formula we apply the LSZ formalism and evaluate the Green's function by employing the effective interaction Lagrangian of eq.(18),

$$\left\langle \overline{N}_a, \vec{p} \,\middle|\, \widetilde{\overline{\pi}}_b(-\vec{p}) \,\middle|\, \overline{N}_c, \vec{0} \right\rangle = (g_1 c_{abc}^1 + g_2 c_{abc}^2) K(\vec{p}) \sqrt{Z_{\pi_b}} P((E_{N_a} - E_{N_c}), E_{\pi_b}), \tag{22}$$

and similar expressions for the other matrix elements. $P(E, \omega)$ denotes again the free particle lattice propagator eq.(7) and $K(\vec{p}) = |\vec{p}|/(4m_N)$ is the appropriate kinematical factor.

The function was fitted to the numerical data using a $\chi^2$ fitting procedure. Masses and energies were taken as input parameters from the 2-point correlation function analysis so that we have 4-fit parameters: two parameters proportional to the coupling constants $g_{NN\pi}$ and $g_{N'N'\pi}$ and two proportional to the mixed matrix elements where the nucleon as well as the $N'$ state contribute. An example for one of the fits is shown in fig.2.

The output of the fit is a product of $Z$ amplitudes, the values of which are also taken from 2–point functions, CG coefficients and coupling constants. To calculate the CG coefficients $c_{abc}^1$ and $c_{abc}^2$, we need to assign $SU(4)_F$ quantum numbers to the states created by the baryon and the pion operators. For the $\pi$ meson we take over the assignment of flavour quantum numbers discussed in the previous section. For the baryon, again using the staggered quark fields $q$, we can find an expression which enables us to investigate the quark content of the operator,

$$N(y) = \frac{1}{3!} \epsilon_{ijk} \sum_A \left( \Gamma_A q_i(y) \Gamma_A^\dagger \right) \times \{ q_j(y) [\mathcal{C} \Gamma_A \otimes \mathcal{C}^* \Gamma_A^*] q_k(y) \}. \tag{23}$$

Here $\mathcal{C}$ denotes the lattice charge conjugation operator and a summation over colour indices $i, j, k$ is understood. By transforming the operator $N(y)$ to momentum space and applying the standard trace theorems for $\gamma$ matrices, it can easily be shown that this operator and the local baryon operator defined in [3] are the same. To find the correspondence between the lattice operator and members of the continuum 20-plet of baryonic $\frac{1}{2}^+$ states, the flavour content of the lattice baryon, which can be read off eq.(23), must be compared with the flavour content of the continuum baryons. This procedure is rather tedious for four quark flavours, leading to the result that the lattice baryon operator corresponds to a linear combination of three continuum states

$$N^{Latt} \leftrightarrow \frac{1}{\sqrt{12}}(\sqrt{3}(A_1^0 + A_2^0) - \sqrt{2}(p_1 + p_2) - (S_1^+ + S_2^+)). \tag{24}$$

The notations for the particle states are taken from ref.[8]. Since in $SU(4)_F$ the 20-dimensional baryonic representation appears twice in the reduction of the representation of a 3-quark state we indicated by lower indices to which of these representations the baryon belongs. Specifically, $S^+$ lies in the 6-dimensional representation of the $u, d, s$ $SU(3)_F$ subgroup of $SU(4)_F$, in a 2-dimensional isospin $SU(2)_F$ representation and has isospin $I_3 = +1/2$. $A^0$ belongs to the $3^*$-dimensional representation of $SU(3)_F$, to a 2–dimensional isospin $SU(2)_F$ representation with isospin $I_3 = -1/2$. $p$ denotes the proton. With the flavour content of the local pion already given in the previous section, the $SU(4)_F$ CG coefficients $c_{abc}^1$ and $c_{abc}^2$ are found to be

$$c_{abc}^1 = -\frac{2 + \sqrt{3}}{36\sqrt{13}}, \tag{25}$$
$$c_{abc}^2 = -\frac{1}{6\sqrt{13}} - \frac{1}{3\sqrt{39}}.$$

To relate the coupling constant of interest $g_{NN\pi}$ to the $SU(4)_F$ couplings $g_1$ and $g_2$, we first consider the case (in the continuum) of a member of the baryon 20-plet belonging to a 2– and a pion state belonging to a 3-dimensional representation of the same $SU(2)_F$ subgroup. Then the strength of the baryon coupling is parameterised by $g_{NN\pi}\hat{c}^{(2)}$ where $\hat{c}^{(2)}$ denotes the corresponding CG coefficient in $SU(2)_F$ for this process. On the other hand, both the baryon and the meson are members of $SU(4)_F$ multiplets and we can alternatively parameterise the coupling by $g_1\hat{c}_1^{(4)} + g_2\hat{c}_2^{(4)}$ according to eq.(18). Since these are just different prescriptions for one and the same physical coupling, we can write

$$g_{NN\pi}\hat{c}^{(2)} = g_1\hat{c}_1^{(4)} + g_2\hat{c}_2^{(4)}. \tag{26}$$

By fitting the numerical data in our simulation to the parameterisation function we extract a number $g$ which is, according to eq.(18), given by $g = g_1 c_{abc}^1 + g_2 c_{abc}^2$ with known $SU(4)_F$ CG coefficients. Using eq.(26) we can relate $g$ to $g_{NN\pi}$ through

$$g_{NN\pi} = g\frac{1}{\hat{c}^{(2)}} \frac{\hat{c}_1^{(4)} + \hat{c}_2^{(4)} g_2/g_1}{c_{abc}^1 + c_{abc}^2 g_2/g_1}, \tag{27}$$

where everything is known except for the ratio $g_2/g_1$. To determine this number we used two different ways.

First, we related the ratio to the experimentally known $F/D$ ratio $g_F^A/g_D^A$ in $SU(3)_F$. This experimental value corresponds to the broken $SU(3)_F$ symmetry between $u$, $d$ and $s$ quarks in the continuum. However, this relation can be regarded as an excellent approximation for the symmetric case. We therefore conclude that the relation between $g_F^A/g_D^A$ and $g_2/g_1$ can be based on the exact $SU(4)_F$ symmetry.

The relation between $g_F^A/g_D^A$ and $g_2/g_1$ can be found by a similar line of arguing which led already to the relation between $g_{NN\pi}$ and $g_1$, $g_2$. We choose baryonic and pionic states and regard them as members of 8–dimensional $SU(3)_F$ representations as well as members of an $SU(4)_F$ 20-plet (baryon) and a 15-plet (pion). The baryon–pion coupling can now be formulated both in terms of $SU(3)_F$ and $SU(4)_F$ couplings and CG coefficients. As the physical result is the same we have

$$(\hat{c}_F^{(3)} g_F^A/g_D^A + \hat{c}_D^{(3)}) g_D^A = (\hat{c}_2^{(4)} g_2/g_1 + \hat{c}_1^{(4)}) g_1, \tag{28}$$

where $\hat{c}^{(3)}$ denote the $SU(3)_F$ and $\hat{c}^{(4)}$ the $SU(4)_F$ CG coefficients. However, we still need the ratio $g_1/g_D^A$ to find the desired relation. Therefore we consider a specific process, e.g. the coupling $\Lambda\Lambda\pi$ where one finds by explicit calculation that $\hat{c}_F^{(3)} = 0$ as well as $\hat{c}_2^{(4)} = 0$. From this we can read off directly the relation

$$g_1 = g_D^A \hat{c}_D^{(3)}/\hat{c}_1^{(4)} = g_D^A / \left( \begin{array}{cc|c} 20 & 15 & 20_1 \\ 8 & 8 & 8_D \end{array} \right) = g_D^A \frac{\sqrt{65}}{4\sqrt{6}}, \tag{29}$$

where the expression in brackets denotes an $SU(3)$ singlet factor [8]. Inserting this result into eq.(28) we find $g_2/g_1 = 1.025$ from the experimental value of $g_F^A/g_D^A = 0.8$ (taken from [9], normalized according to de Swarts phase conventions for $SU(3)$ Clebsch-Gordan coefficients [10]). This in turn enables us to find the desired relation between $g$ and $g_{NN\pi}$:

$$g_{NN\pi} = 1.6891 g. \tag{30}$$

Based on these ingredients, we extract a value for $g_{NN\pi}$ from our numerical data. We find

$$g_{NN\pi} = 13.8 \pm 5.8, \tag{31}$$

which is to be compared with the experimental value of $13.13 \pm 0.07$. The result is consistent with the value quoted in ref.[5] $g_{NN\pi} = 12.7 \pm 2.4$ which was obtained by using Wilson fermions.

The second method is based solely on lattice calculations. For this purpose we numerically analysed a second correlation function using the same baryonic operator but a different pion $\pi'$. This leads to a set of different $SU(4)_F$ CG coefficients $\tilde{c}_{abc}^1$ and $\tilde{c}_{abc}^2$ and a different coupling $g' = g_1 \tilde{c}_{abc}^1 + g_2 \tilde{c}_{abc}^2$ so that we get two independent relations for the $SU(4)_F$ couplings. Under the assumption of flavour symmetry $g_1$ and $g_2$ have the same value for all pion states in the 15-plet and we are able to solve for $g_1$ and $g_2$ separately.

We used the spatially non-local pion operator [No.14] for this purpose which corresponds to the continuum operator

$$\pi' \leftrightarrow \frac{1}{\sqrt{2}} \left( \frac{\sqrt{6}}{3} \eta_{15} - \frac{1}{\sqrt{3}} \eta_8 - \pi_0 \right), \tag{32}$$

from which we derive the CG coefficients

$$\tilde{c}^1_{abc} = \frac{61(2+\sqrt{3})}{576\sqrt{13}}, \tag{33}$$

$$\tilde{c}^2_{abc} = -\frac{5(3+2\sqrt{3})}{18\sqrt{13}}.$$

If we use the values for $g$ and $g'$ extracted from our numerical data, we are able to compute $g_2/g_1$,

$$\frac{g_2}{g_1} = 0.7(4). \tag{34}$$

The large uncertainty of $g_2/g_1$ within the second method is mainly due to statistical noise in the numerical data. A possible explanation for the poorer quality of the $NN\pi'$ 3-point function compared to the $NN\pi$ correlation function, is the non-local character of the $\pi'$ operator which requires the insertion of gauge fields in order to ensure local gauge invariance.

The result for the ratio quoted above is smaller than the value $g_2/g_1 = 1.025$ we obtained previously, though both are compatible within the errors. But we have also to bear in mind that the second value was obtained under the assumption of $SU(4)_F$ flavour symmetry on the lattice. However, the discrepancy between the masses of the lattice Goldstone pion and e.g. the mass of $\pi'$ we found in our spectrum analysis indicates that $SU(4)_F$ flavour symmetry is not yet completely restored for our lattice parameters. In order to roughly estimate the effect of flavour symmetry breaking, one could argue that, as a result of the higher mass of the $\pi'$ state, the virtual transition $N \to N + \pi'$ is suppressed, thus leading to a smaller value for $g'$. If $g'$ increases, the ratio $g_2/g_1$ rises and comes closer to 1.025.

## Acknowledgement

This work was supported in part by the Deutsche Forschungsgemeinschaft. The numerical computations were performed on the Cray Y-MP in Jülich, with time granted by the Scientific Council of the HLRZ as part of the MTc project. We would like to thank both institutions for their support.

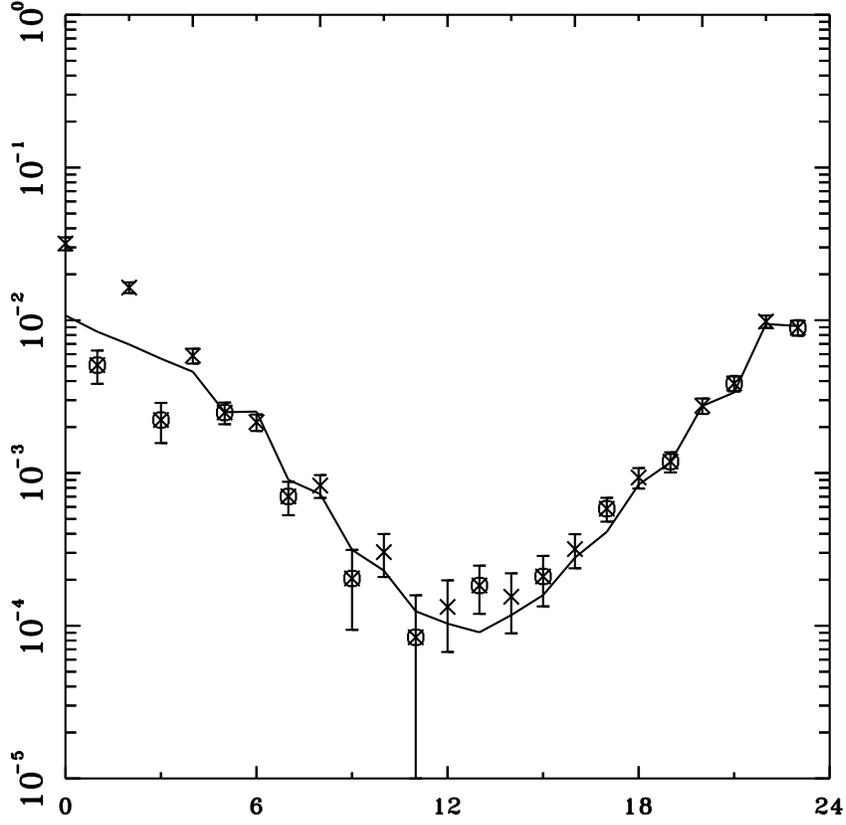

Figure 1: The numerical data for the 3–point function in eq.(2) with the local $\rho$ operator (positive values are marked with crosses, negative values with squares). A fit from $t_1 = 1$ to $t_1 = 23$ without $t_1 = 4$ (the time slice of the pion source) is shown by the full curve.

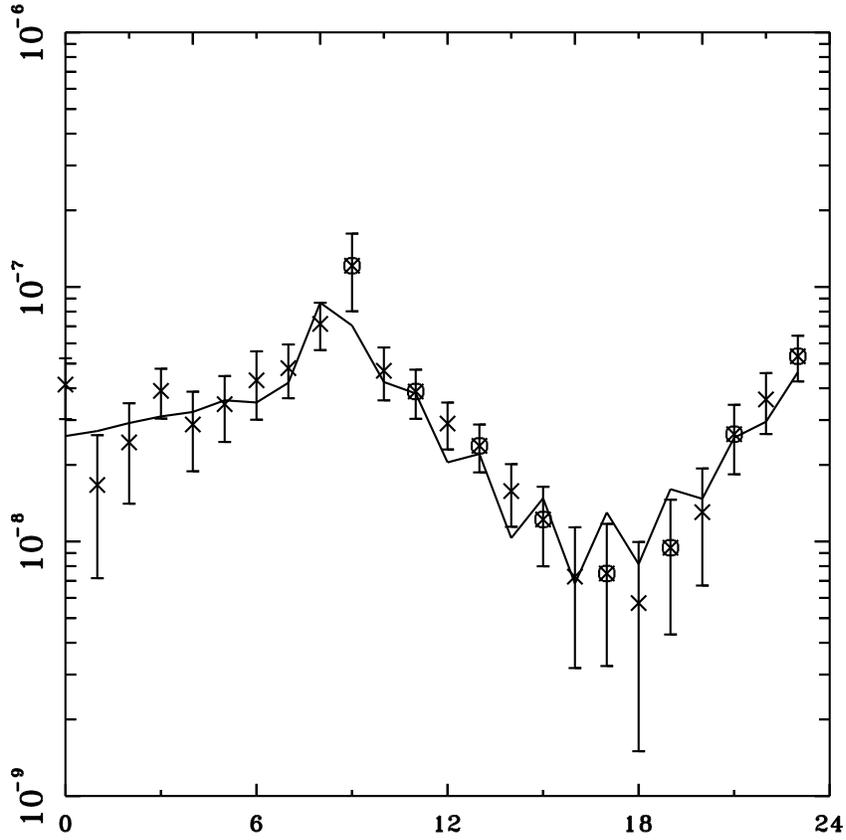

Figure 2: The numerical data for the 3–point function defined in eq.(19) with the local Goldstone $\pi$ operator (positive values are marked with crosses, negative values with squares). Also shown is a fit from $t_2 = 1$ to $t_2 = 23$ without $t_2 = 8$ (the time slice of the nucleon).

14